\documentstyle{crnart12}
\def\beq{\begin{equation}}
\def\eeq{\end{equation}}
\def\bea{\begin{eqnarray}}
\def\eea{\end{eqnarray}}
\def\bq{\begin{quote}}
\def\eq{\end{quote}}

\def\gappeq{\mathrel{\rlap {\raise.5ex\hbox{$>$}}
{\lower.5ex\hbox{$\sim$}}}}

\def\lappeq{\mathrel{\rlap{\raise.5ex\hbox{$<$}}
{\lower.5ex\hbox{$\sim$}}}}

\begin{document}
\pagestyle{empty}
\begin{flushright}
{CERN-TH.6805/93}
\end{flushright}
\vspace*{5mm}
\begin{center}
{\bf Could a nearby supernova explosion have caused a mass extinction?} \\
\vspace*{1cm} {\bf John Ellis} \\
\vspace*{0.3cm}
Theoretical Physics Division, CERN \\
CH - 1211 Geneva 23 \\
\vspace*{0.5cm}
and \\
\vspace*{0.5cm}
{\bf David N. Schramm} \\
\vspace{0.3cm}
University of Chicago, Chicago, IL 60637, U.S.A. \\
and \\
NASA/Fermilab Astrophysics Group, \\
Batavia, IL 60510, U.S.A. \\
\vspace*{2cm}
{\bf ABSTRACT} \\ \end{center}
\vspace*{5mm}
\noindent
We examine the possibility that a nearby supernova explosion could have
caused one or more of the mass extinctions identified by
palaeontologists. We
discuss the likely rate of such events in the light of the recent
identification of Geminga as a supernova remnant less than 100 pc away
and the discovery of a millisecond pulsar about 150 pc away,
and            observations of SN 1987A. The fluxes of $\gamma$ radiation
and charged cosmic rays on the Earth are estimated, and their effects on
the Earth's ozone layer discussed.
A supernova explosion of the order of 10
pc away could be expected every few hundred million years, and could
destroy the ozone layer for hundreds of years, letting in
potentially lethal solar ultraviolet radiation.
In addition to
effects on land ecology, this could entail mass destruction of plankton
and reef communities, with disastrous consequences for marine life as
well. A supernova extinction should be distinguishable
from a meteorite impact such as
the one that presumably killed the dinosaurs.

\vspace*{3cm}


\begin{flushleft} CERN-TH.6805/93 \\
February 1993
\end{flushleft}
\vfill\eject

\setcounter{page}{1}
\pagestyle{plain}

During the 600 million years or so since life on Earth emerged from its
murky  pre-Cambrian beginnings, it has been subjected to five major mass
extinctions, the ''Big Five", as well as a spectrum of
lesser extinctions \cite{aaa}.
These have been the subject of intensive research, particularly during the
last decade. Many theories  have been advanced to explain one or more of
these extinctions, including both terrestrial and astrophysical events.
Among the former, one should mention massive volcanic episodes. Among the
latter, particular mention should be made of meteorite impacts, whose
advocacy by Alvarez et al. \cite{bb} stimulated much research.
The famous mass extinction
at the end of the Cretaceous, which finished off the dinosaurs,
has been convincingly identified with such a meteorite impact, whilst the
record-holding Permian extinction might have been caused by the volcanic
episode that created the Siberian traps.
Advocacy of these volcanic and
meteoritic mechanisms has been aided by the availability and tangibility
of supporting evidence in the forms of large lava flows and contemporary
volcanoes on the one hand, and impact craters and Earth-crossing
asteroids on the other hand.

However, these are not the only mechanisms to have been proposed. Among
astrophysical origins of mass extinctions, we mention variations in the
solar constant,  supernova explosions, and meteorite or comet impacts that
could
due to perturbations of the Oort cloud. The first of these has little
experiment
support. Nemesis \cite{dd}, a conjectured binary companion of the Sun, seems to
been excluded as a mechanism for the third \footnote{Except
possibly for a small region of parameter space \cite{VS}.},
although other possibilities such as
passage of the solar system through the galactic plane may still be tenable.
The
supernova mechanism \cite{ee} has attracted less research interest than some of
others, perhaps because there has not been a recent supernova explosion in our
g
concentrate our minds, and perhaps because the prospective lethality of a
nearby
supernova explosion has not been fully appreciated.

We think that there are at least four reasons for reconsidering now the
supernova mechanism for mass extinction. One is that extinction studies
have advanced greatly since the supernova mechanism was last discussed.
Another is that the identification \cite{ff} of the Geminga $\gamma$ source
with
supernova remnant about 60 pc away, that apparently exploded about
300,000 years ago, confirms that such nearby events are not fanciful and
provides us with new hints about rates,
as does the recent discovery of a millisecond pulsar
PSR J0437-4715 about 150 pc away.
A third reason is that the recent
detailed observations of SN 1987A clarify the characteristics of supernova
explosions \cite{ggg}. Finally, there has been much recent
work on the biological
effects of ozone depletion, motivated by the observed Antarctic
hole \cite{hh}.
It was
Ruderman \cite{jj} who first pointed out the possible effect
of a supernova on the
ozone layer, and this seems to us potentially the most
catastrophic effect
of a nearby supernova explosion.

The best supernova rate estimate we can offer indicates that one or
more supernova explosions are likely to have occurred within 10 pc
 or so
of the Earth during the Phanerozoic era, i.e.,
during the last 570 million
years since the sudden biological diversification at the start of the
Cambrian. Since stars' orbital motions around the galaxy can separate
them by up to 10 Kpc over 100 million years, the remnants of any such
supernova explosions would not be very
close  today. On the
other hand, the space within 10 pc or so of Earth should contain
remnants of explosions that took place within the last 100 million years
up to 10 Kpc away. The best estimate we can offer of the fluxes of
energetic electromagnetic and charged cosmic radiation from a supernova
explosion within 10 pc indicates that the latter would have destroyed
the Earth's ozone layer over a period of $\sim$ 300 years or so. Recent
studies,
motivated by the appearance \cite{hh} of the ozone
depletion in the Antarctic,
indicate that the increase in
ultraviolet radiation due to
ozone removal could have a negative effect \cite{kk} on phytoplankton,
and hence on
the rest of marine life, from zooplankton through to benthic life, in
addition to the obvious threat to terrestrial life.
Since reef communities
are also dependent on photosynthesizing organisms, they could also have
been severely damaged or disrupted by the ozone hiatus, with
correspondingly severe consequences for the diverse marine life they
support. We also note that a shutdown of photosynthesis
due to solar ultraviolet
irradiation could well be followed by a greenhouse episode
\footnote{Possible damage to DNA is also a cause for concern
\cite{DLA}.}.

We first discuss the likely rate of supernova explosions in the light of
Geminga and other recent developments. Many authors have estimated that
there are explosions every 10 to 100 years in our galaxy, which contains
about $10^{11}$ stars. We draw the reader's attention in particular to a
recent analysis \cite{lll} of the amount of oxygen in the galaxy, which
originat
from supernova ejecta and seems to require an average explosion rate of about
on
10 years, if all the ejecta are retained in the galaxy. However, the local hole
apparently due to Geminga extends much further into the less dense region
 away from our
local spiral arm, raising the possibility that material ejected out of the
galac
plane might escape altogether. In this case, explosions at average
intervals even shorter than 10 years could be required, despite their
observational rarity in other galaxies. This is conventionally
 ascribed to obscuration, but could also be due to the same reason that SN
1987A
was relatively dim, namely the previous loss of its outer envelope.
SN 1987A would not have been seen in most surveys if it had
occurred in a distant
galaxy. Taking an average stellar density of 1 pc$^{-3}$,
a supernova rate of 0.1
y$^{-1}$ corresponds to one explosion every 240 million years on average
within 10 pc of the Earth. Some might consider this rate optimistic (or
pessimistic, depending on one's point of view), but scaling from Geminga
(albeit with a statistics of one!) suggests an even larger rate:
assuming a
distance of 60 pc and an age of 300,000 years inferred from the rate of
deceleration of Geminga's spin~\footnote{It would be interesting
to consider whether any trace of the Geminga explosion could be
found as an isotope anomaly in ancient
ice: specific predictions should be explored.},
one finds an explosion within 10 pc
every 70 million years or so.
A relatively high rate is also indicated by the recent
discovery \cite{SJA} in a partial sky survey
of the millisecond pulsar PSR J0437-4715 about
150 pc away, with a spin-down age of about 10$^9$ years.
Assuming that we are in its beam cone, and that this
subtends about $10^{-2}$ to $10^{-3}$ of the full solid
angle, simple scaling indicates that a supernova explosion
could occur within 10 pc of us every 500 My or so.
Inferring supernova rates from pulsars is known to be
quite uncertain, but we feel the consistency is nonetheless
interesting.
We conclude that it is very plausible
that there have been one or more supernova explosions within 10 pc of
the Earth during the Phanerozoic era.

Three more comments on galactic supernovae might be useful. One is that
they mainly occur in the spiral arms of the galaxy, so that the rate
should not be expected to be uniform in time. The Earth passes through a
spiral arm once every 100 million years or so, with each passage taking about
10
million years, though it is unclear whether this would lead to
any discernible
periodicity in nearby supernova events. In any case, this period does not
coinci
with the reported $\sim$ 26 to 30 million-year periodicity of the bulk of
extinc
events: anyway, we do not
expect supernova extinctions
to constitute the bulk of the known extinctions \cite{aaa}.
A second comment is that the
relative velocities of stars in the galaxy mix them up very thoroughly on a
time
of about 100 million years: for example, Geminga's proper motion corresponds to
transverse velocity of about 30 km/s, sufficient to take it 10 Kpc away from us
the next 100 million years. This means that the remnants of any nearby
explosion
be far away by now. It also means that no star now in the solar neighbourhood
is
obvious threat to our survival. Thirdly, if we are right, the solar
neighbourhoo
be populated  with remnants of explosions that took place long ago and far
away,
would be interesting to devise an observational programme to scan
for them,
perhaps in X-ray or radio bands, as a check on our proposal.

We now present some crude estimates of
the likely terrestrial effects of a nearby supernova
explosion. Because of the simple $1/R^2$ scaling law for intensity, it is
genera
agreed that the heating of the Earth would not be significant, and that the
opti
brightness of a supernova at 10 pc would not greatly harm the ecology. It is
als
to convince oneself that supernova ejecta would not have a significant effect
on
apparent solar constant. The most important effects are likely to be
those of ionizing radiation, which falls into two categories. There is a
burst of neutral electromagnetic radiation that arrives over a period of
a few months, and a larger and longer burst of charged cosmic ray
particles. In line with previous estimates
\cite{hh},\cite{KDT},\cite{HLA},
we assume that the neutral
component has a total energy output of $3.10^{46}$ ergs, and the
charged component $10^{50}$ ergs. The period over which the latter are
emitted is unclear \footnote{Models of cosmic ray acceleration
predict ranges from 10 to 10$^5$ years or more \cite{be}.},
but the charged cosmic ray burst is in any case
spread out by diffusion through the inhomogeneous galactic magnetic
field. Taking an angular persistence length of 1 pc for the latter,
one estimates a diffusion time of $3.{D_{\rm pc}}^2$ years
\cite{HLA}
where $D_{\rm pc}$
is the distance of the supernova measured in pc. The average flux of neutral
ionizing radiation per unit surface area normal to the Earth's surface is
theref
estimated to be
\beq
\phi_n = \frac{3 \times 10^{46}}{16\pi D^2}~{\rm ergs/cm}^2 \simeq
6.6 \times 10^5 \left( \frac{10}{D_{pc}} \right)^2~{\rm ergs/cm}^2
\label{1}
\eeq
for about a year, whereas the average normal flux of charged cosmic rays is
estimated to be
\beq
\phi_{cr} = \frac{10^{50}}{16\pi
D^2(3D^2_{pc})}~{\rm ergs/cm}^2
\cdot {\rm y} \simeq 7.4 \times
10^6~\left( \frac {10}{D_{pc}}\right)^4~{\rm
ergs/cm}^2 \cdot {\rm y}
\label{2}
\eeq
  for a duration
of about $3.{D_{\rm pc}}^2$y. For comparison, the ambient cosmic ray flux is
 $9.10^{4}$
ergs/cm$^2$/y, which produces a radiation dose at the Earth's surface of 0.03
R/
$10^7 NO$ molecules/cm$^2$/y after diffusion throughout the stratosphere. We
dou
an increase in the cosmic-ray-induced radiation dose by one or two orders of
mag
as suggested by the numbers (1,2) for a supernova 10 pc away, would be
catastrop
the global ecology, though we cannot exclude the possibility that it would be
ha
to some key organisms. However, we do believe that a dramatic increase in $NO$
production would have catastrophic implications for the Earth's ozone layer,
and
for many life-forms.

We use the analysis of Ruderman \cite{jj}, who was the first to consider the
effect of a supernova explosion on the Earth's ozone layer, to estimate
the increase in $NO$ production and the consequent ozone destruction.
Ionizing radiation is estimated to produce $NO$ at a rate of about
\beq
R_{NO} = 9 \times 10^{14}~\left(\frac{\phi}{9
\times 10^4} \right) \times \left( \frac{13}{10+y}
\right)~{\rm molecules/cm}^2 \cdot {\rm y}
\label{3}
\eeq
 if $NO$ dominates over $NO_2$ in the stratosphere, as we expect,
where $y$ is the $NO$ abundance in parts per $10^9$.  The first factor in
(3) is the present rate of $NO$ production by cosmic rays, the next is the
ratio
supernova radiation to cosmic radiation, and the last is a ratio of
efficiencies
assuming a present $NO$ abundance of 3 parts per $10^9$.  We assume for
simplici
the electromagnetic and cosmic radiation from a supernova ionize at the same
rat
erg of incident energy as do present-day cosmic rays. We therefore expect
that the charged cosmic radiation from the supernova would produce
significantly
$NO$ than would the electromagnetic radiation, in an amount
\beq
R_{NO} = 7.4 \times 10^{16}~\left( \frac{10pc}{D}
\right)^4~\left( \frac{13}{10+y} \right)~{\rm
molecules/cm}^2 \cdot{\rm y}
\label{4}
 \eeq
during about $3D^2_{pc}$ years.  The residence time for $NO$ in the
stratosphere
diffusing out is thought to be 2 to 6 years. Taking a mean of 4 years, and
dividing $R_{NO}$ by the stratospheric column density of $5.10^{23}
molecules/sq.cm$, we find that the supernova cosmic rays would contribute
\beq
y_{cr} = 600 \left( \frac{13}{13+y_{cr}}
\right)~\left( \frac{10}{D_{pc}} \right)^4
\simeq 88~\left( \frac{10~pc}{D}
\right)^2
\label{5} \eeq
to the $NO$ abundance in parts per $10^9$. Assuming an
altitude-independent abundance of $NO$, Ruderman gave the following approximate
for the ratio of $O_3$ to present ambient $O_{3_0}$:
\beq
F_0 \equiv \frac{O_3}{O_{3_0}} = \frac {\sqrt{16+9X^2} - 3X}{2}
\label{6}
\eeq
where $(3+y_{cr})/3$ is the ratio of $NO_0$ to present
ambient $NO$.  Equation (6) may be approximated by $\frac{4}{3X} \simeq
\frac{4}{y_{cr}}$ for large $X(y_{cr})$.

The resulting increase in the penetrating flux of solar ultraviolet
radiation, integrated over the duration of the cosmic ray burst, is
\beq
(f^{F_0} - f) \times (3D^2_{pc})
\label{7}
\eeq
where $f$ is the fraction of the incident solar ultra-violet radiation
that normally reaches the Earth's surface. In the case of radiation with
a wavelength of $2500 A$, which has the maximum relative effectiveness
for killing $Escherichia~ coli$ bacteria and a high relative efficiency
for producing erythema (sunburn), $f$ is about $10^{-40}$ today, so a
reduction of the $O_3$ layer to 10\% of its present thickness would
increase the flux of ultraviolet radiation by 36 orders of
magnitude. For nearby supernovae, the integrated increase in the
penetrating flux can be approximated by
\beq
10^{-\left( \frac{D}{7~pc} \right)^2}
\times
300~\left( \frac{D}{10~pc}\right)^2
\label{8}
\eeq
 We see that the flux increase is probably negligible for supernovae
much more than 10 pc away, whilst their rate is probably negligible for
supernovae much closer than 10 pc. Hence our focus on 10 pc as the
critical distance around which a supernova explosion is most likely to
have caused a mass extinction, which we have taken as a reference in
our flux estimates above.

A species may become extinct either because it is killed directly, for
example by  sunburn or a radiation overdose,
or for some indirect reason, for example
a change in the environment, such as global cooling or warming, or the
disruption of its food supply. A nearby supernova explosion could affect
many species directly via the solar ultraviolet radiation admitted after
destruction of the ozone layer. These would need to be studied on a
case-by-case basis. Apart from this increase in radiation,
we do not expect
any dramatic environmental effects resembling those caused by a large
meteorite impact or massive vulcanism. Instead, we focus here on the
possibility of mass extinction caused by a disruption of the food
chain at a low level, specifically by the destruction of
photosynthesizing organisms. This has already been discussed as an
important side-effect of a large impact or volcanic episode. In our case,
it is clear that any photosynthesizing organism must try to ``see" the
Sun, and the absence of an ozone layer means that it will ``see" and be
affected by ultraviolet radiation as well. Photosynthesis manifests both
a diurnal and an annual cycle. An ozone hole induced by a supernova
explosion 10~pc away would last for about 300 years
(to within an order of magnitude), and hence act over
many annual cycles, indeed, longer than the lifetimes of most
present-day fauna.

Half of photosynthesis today is due to phytoplankton, and the effect on
them of ultraviolet radiation has recently been
studied in connection with
the ozone hole in the Antarctic \footnote{As we have already noted,
damage
to DNA is also a cause for concern \cite{DLA}.}.
A decline in the rate of photosynthesis by
of Antarctic
plankton exposed in
plastic bags has been demonstrated \cite{kk}. The possible importance
of radiation effects on polyethylene as a factor in this particular
experiment has
been emphasized \cite{HHH}, but these objections are not seen as
conclusive \cite{ps}. Therefore, we feel that this
experiment makes an {\it a~ priori} case that a long-term exposure
to the  full
ultraviolet radiation of the Sun could shut down marine photosynthesis,
and hence cause a mass extinction of marine life, from phytoplankton to
zooplankton and so on all the way to benthic organisms.
We note that a shutdown of
photosynthesis due to ultraviolet irradiation from the
Sun could lead indirectly to a
greenhouse episode, due to a build-up of CO$_2$.
We note also that reef communities,
which are known to have been destroyed during mass extinctions,
are particularly
exposed to solar ultraviolet radiation and depend directly on
photosynthesizing
organisms, and we
remind the reader that reef communities are the
source of much of the
marine biodiversity. Thus the effects of a nearby
supernova explosion would
not be limited to terrestrial organisms, and might even have had a
larger effect on the marine community. Could such an event have been
responsible for the Permian
mass extinction which finally killed the
trilobites?

We conclude that recent observations of Geminga,
PSR J0437-4715 and
SN 1987A strengthen the case for one
or more supernova extinctions during the Phanerozoic era.
A nearby supernova
explosion would have depleted the ozone layer, exposing
both marine and terrestrial
organisms to potentially lethal solar ultraviolet radiation.
In particular,
photosynthesizing organisms including phytoplankton and reef
communities are likely
to have been badly affected. We believe that the potential
signatures of supernova
extinctions merit further study.

\vspace*{2cm}
\noindent
{\bf Acknowledgements}

We thank the organizers of the Texas/PASCOS Symposium in Berkeley in
December 1992 for stimulating our interest in this topic, and Neil
Gehrels, Stan Woosley, Peter~Vandervoort,
David Raup,    John Frederick and Douglas Morrison
for useful
discussions. J.E. thanks LAPP for hospitality while some of this
work was done, and the work of D.N.S. was supported in part by the
US NSF, NASA and the Department of Energy at the University of
Chicago, and NASA and the Department of Energy at Fermilab.

\vfill\eject


\begin{thebibliography}{99}
\bibitem{aaa} D. Raup, 1991, {\it Extinction:
Bad Genes or Bad Luck}  (W. Norton, New
York) and references therein.
\bibitem{bb} L. Alvarez, W. Alvarez, F. Asaro and H. Michel, 1980,
{\it Science} {\bf 208}, 1095-1108.
\bibitem{dd} M. Davis, P. Hut and R. Muller, 1984, {\it Nature} {\bf 308}
, 715-717.
\bibitem{VS} P.O. Vandervoort and E. Sather, 1993, U. of Chicago
preprint, {\it Icarus} (in press).
\bibitem{ee}D. Milne, D. Raup, J. Billingham,
K. Niklaus and K. Padien, 1985, {\it The Evolution of Complex
and Higher Organisms}, NASA Report SP-478.
\bibitem{ff} N. Gehrels and W. Chen, 1993, {\it Nature}, in press.
\bibitem{ggg} W.D. Arnett, J. Bahcall, R. Kirgher and S. Woosley, 1989, {\it
Ann.Rev.Astron. \& Astrophys.} {\bf 27}, 629.
\bibitem{hh} X. Niu et al., 1992, {\it J.Geophys.Res.} {\bf 97}, 14661.
\bibitem{jj} M.A. Ruderman, 1974, {\it Science} {\bf 184}, 1079-1081.
\bibitem{kk} R.C. Smith et al., 1992, {\it Science} {\bf 255}, 952.
\bibitem{DLA} D. Lubin et al., 1992, {\it J.Geophys.Res.}
{\bf 97}, 7817-7828.
\bibitem{lll} W. Arnett, D. Schramm and J. Truran, 1989,
{\it Ap.J.} {\bf 339}, L25-L27.
\bibitem{SJA} S. Johnston et al., 1993, {\it Nature}
{\bf 361}, 613-615.
\bibitem{KDT} K.D. Terry and W.H. Tucker, 1968, {\it Science}
{\bf 159}, 421-423;\\
W.H. Tucker and K.D. Terry, 1068, {\it Science} {\bf 160},
1138-1139.
\bibitem{HLA} H. Laster, 1968, {\it Science} {\bf 160}, 1138.
\bibitem{be} R. Blandford and D. Eichler, 1987, {\it Phys. Rep.}
{\bf 154}, 1-75.
\bibitem{HHH} O. Holm-Hansen and E.W. Hebling, 1993, {\it Science}
{\bf 259}, 534.
\bibitem{ps} B.B. Prezelin and R.C. Smith, 1993, {\it Science}
{\bf 259}, 534-535.
\end{thebibliography}
\end{document}